\documentclass[aps,prb,twocolumn,groupedaddress,showpacs]{revtex4-1}

\usepackage{graphicx}% Include figure files
\usepackage{dcolumn}% Align table columns on decimal point
\usepackage{bm}% bold math

\begin{document}

%\preprint{}

\title{Stochastic Pumping of a Polariton Fluid}


\author{M.~A\ss mann}%
\author{M.~Bayer}%
\affiliation{%
Experimentelle Physik 2, Technische Universit\"at Dortmund, 44221
Dortmund, Germany
}%
%\thanks{}


\date{\today}

\begin{abstract}
We investigate the response of a polariton laser driven slightly off-resonantly using light fields differing from the routinely studied coherent pump sources. The response to driving light fields with thermal and displaced thermal statistics with varying correlation times shows significant differences both in the transmitted intensity, its noise and the position of the non-linear threshold. We predict that adding more photons on average may actually reduce the transmission through the polariton system. 
\end{abstract}

%\keywords{}

\maketitle
\section{Introduction}
In optical spectroscopy, most experiments are performed with coherent light fields due to their appealing properties. They can be created with well defined and easily tunable average photon number, have a well-defined phase and as eigenstates of the photon annihilation operator they are insensitive to loss and non-ideal detection efficiency. Nevertheless, recently it has been suggested that additional photon number noise may indeed prove beneficial for example in experiments based on multi-photon absorption \cite{Jechow2013} or stochastic resonance \cite{Abbaspour2014}. Further, it has been demonstrated that tailoring the photon statistics of an excitation light field may provide significantly deeper insights into the response and dynamics of a system \cite{Assmann2011a, Berger2014} or can be used to distinguish different many-body effects \cite{Kira2011,Hunter2014}.\\Here, we study the interaction of light with thermal photon number statistics with a polariton fluid in a microcavity. Exciton-polaritons are quasiparticles consisting of excitons and photons in the strong coupling regime. The Coulomb interaction between them arising due to the excitonic part results in significant interparticle scattering. The photonic part results in a very light mass of polaritons. As the photonic component is a part of the polariton wave function, polaritons couple to light fields. Therefore, they can be excited optically and the light emitted from a polariton system directly represents its state. One of the most intriguing properties a polariton fluid may exhibit is bistability. When pumped at normal incidence, the transmission of an incident light beam depends on the spectral overlap between that light field and the polariton mode. The latter depends on the occupation of the polariton system. Due to the excitonic polariton component, interactions between polaritons of the same spin result in a blueshift of the polariton state. If the pump light field itself is at higher energy compared to the polariton ground state, these interactions will increase the spectral overlap between these two modes, which will in turn lead to the excitation of more polaritons. At some pump photon number $n_{on}$, this feedback loop gives rise to a strongly non-linear behavior and the system turns into the on-state, yielding high transmission. When reducing the pump intensity again, the system will transit back into the off-state again, but the threshold $n_{off}$ will be lower than $n_{on}$. The region between these two thresholds is the bistability regime, where two solutions for the transmission of the system exist, depending on the occupation number of the polariton system.

\section{Polariton model and pump photon statistics}
This simple picture is valid for a pump density that is well described by its mean value and only subject to small fluctuations. That prerequisite is fulfilled for coherent laser light fields, which show a Poissonian photon number distribution around the mean value
\begin{equation}
p_{coh}(n)=\frac{\langle n \rangle^n e^{-\langle n \rangle}}{n!},
\end{equation}
but not for a wide range of important other light sources like thermal light emitted from a blackbody. Thermal light follows a Bose-Einstein photon number distribution
\begin{equation}
p_{th}(n)=\frac{\langle n \rangle^n}{(1+\langle n \rangle)^{n+1}},
\end{equation}
for which zero is always the most probable photon number and the photon number distribution is very broad. This implies that when pumping a polariton fluid using a thermal light field,even if the mean pump photon number is above $n_{on}$, there will still be times when the instantaneous pump photon number is below $n_{off}$. Depending on the timescale of the photon number fluctuations, the polariton fluid may actually never reach a true steady state, but move back and forth between the on and off states in a random manner. A reasonable model of the dynamics of a driven polariton fluid $\psi$ is given by the following equation:
\begin{equation}
\frac{d \psi}{d t}=(-\frac{1}{2}\gamma_c-\frac{i}{\hbar}(\Delta+g \left|\psi\right|^2))\psi + P(t),
\end{equation}
where $P$ represents the time-dependent amplitude of the pump field, $\gamma_c$ denotes the cavity decay rate, $\Delta$ represents the energy detuning between the pump energy and the polariton ground state and $g$ is the polariton-polariton interaction constant \cite{Baas2004}. For simplicity, we assumed circularly polarized pumping, so that only one spin species is present. Due to the spin-anisotropic polariton-polariton interaction, a much more complex behavior is expected when linearly polarized pumping is used or polariton spin flip processes become important \cite{Gippius2007}. For our simulations we used values of $\gamma_c$=0.1\,ps$^{-1}$, $g$=6\,$\mu$eV and $\Delta$=-2\,meV.\\Typically simulations use a constant driving term. The case of coherent pumping with added coloured noise has also been investigated \cite{Maslova2009}. In order to evaluate the differences between a classical coherent pump and thermal driving, we simulate the thermal light in terms of a random walk of $n$ independent oscillators \cite{Loudon1973}.
\begin{figure}[t]
\includegraphics[width=\linewidth]{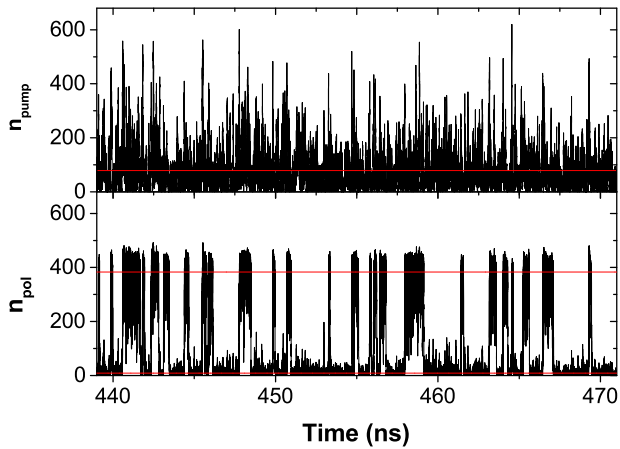}
\caption{(Color online) Comparison of the two pumping schemes. Upper panel: Thermal (black line) and coherent (red line) pump photon number $n_{pump}$ time streams for the same mean photon number $\langle n\rangle_{pump}$=79. Bottom panel: Time series of the polariton occupation number $n_{pol}$. For the coherent pump beam, the system may be in the on-state (upper red line) or the off-state (lower red line) depending on the initial polariton number. For the thermal pump (black line), the system state switches irregularly between the on- and off-states.} \label{fig:TimeSeries}
\end{figure} 
Each oscillator is initialized with a random phase and performs a random phase jump per time bin. By setting the maximum phase change each oscillator can undergo per simulation time bin, we can directly control the correlation time of the light field. We simulate a total time window of 1\,$\mu$s and average over 20 realizations of the random thermal pump field.\\Figure \ref{fig:TimeSeries} demonstrates the central difference between the two pumping schemes in terms of a time stream of the pump photon numbers at the same mean pump density and the corresponding polariton occupation numbers. For coherent excitation (red lines) the mean pump photon number is constant and depending on the initial state of the cavity, the polariton system will be in the on- or off-state constantly. For thermal excitation (black lines), the pump photon number is fluctuating strongly and the polariton system repeatedly switches back and forth between the on- and off-states. Here, the long-term behavior does not depend on the initial condition of the cavity. The results show that the system is likely to switch to the on-state after short pump intensity bursts. The average time the system spends in the on-state is significantly longer than the duration of the pump intensity bursts. Thus, it seems natural to assume that the average on-time depends more strongly on the cavity parameters than on the pump beam properties, while the process of switching to the on-state is clearly determined by the pump field. Accordingly, it seems worthwhile to study not only the influence of the magnitude of the pump photon number noise, but also the typical time scale of the fluctuations, which will be called the correlation time $\tau_c$ in the following.

\section{Polariton Occupation Number}
\begin{figure}[t]
\includegraphics[width=\linewidth]{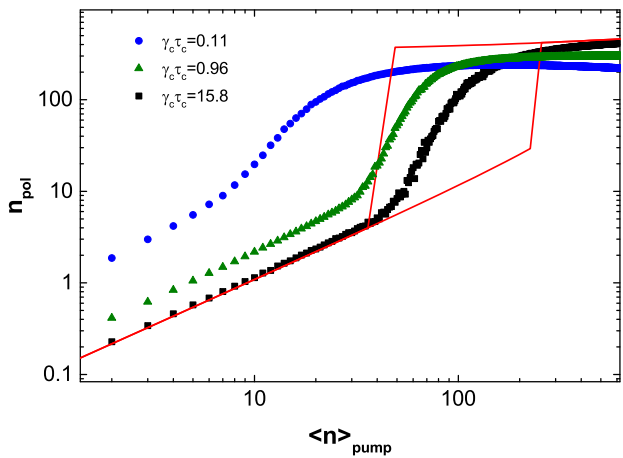}
\caption{(Color online) Mean polariton occupation number for pumping with thermal light fields with different $\tau_c$. The solid red line denotes the corresponding curve for coherent pumping.} \label{fig:CohTime}
\end{figure}

The most basic quantity which may show a dependence on photon statistics is the mean polariton occupation number or equivalently the transmission through the cavity. Figure \ref{fig:CohTime} shows a direct comparison of the polariton fluid mean occupation numbers under coherent and thermal pumping. For a coherent pump source, the bistability region can be seen clearly, while there is only one curve for thermal pumping. In the latter case, the shape of the transmission curve shows a peculiar dependence on the pump field correlation time. For slowly varying thermal pump fields and low pump photon numbers below the bistable region, there is almost no difference between coherent and thermal pump fields. In this regime, the transmission through the cavity is mainly governed by the mean photon number of the pump light field. Within the bistable region, the thermal transmission curve follows an intermediate shape between the on- and off-state seen for coherent pumping. At the lower end of the bistable region, the thermal response is closer to the off-state, while it approaches the on-state at its upper end. For pump photon numbers above $n_{on}$, the thermal response almost equals the coherent response again.
For shorter correlation times and, accordingly, faster fluctuations, the results show a remarkable difference. Below the bistability threshold, the transmission is enhanced significantly and proportionally to the inverse correlation time. The onset of the strong non-linear increase in the transmission curve is reached at much lower $\langle n\rangle_{pump}$, which may be up to one order of magnitude lower compared to the case of coherent pumping. Surprisingly, the transmission reaches a maximum within the bistable region and decays slowly afterwards. These results can be readily understood by taking the different time scales of the photon number fluctuations and the cavity decay rate into account. Due to the fluctuations present, $n_{pump}$ will sometimes exceed $n_{on}$ and the polariton system will switch to the on-state. At some point, $n_{pump}$ will fall below $n_{off}$ and the system will develop back to the off-state again. However, the crucial quantity that determines whether the system is in the on- or off-state is the spectral overlap between the pump beam and the polariton mode, which is governed by $n_{pol}$ and not by $n_{pump}$. If $n_{pump}$ drops again, the corresponding decay of $n_{pol}$ will occur with a delay governed by the cavity lifetime, which is proportional to $\gamma_c^{-1}$. If the fluctuations in $n_{pump}$ are slow, this delay is negligible and $n_{pol}$ will follow $n_{pump}$ almost instantaneously. The cavity transmission reduces to a function of $n_{pump}$. For faster fluctuations, the delay may become comparable to the time scale of the fluctuations. In this case, the photon numbers will return to values below $n_{off}$, but the slower cavity decay time will keep the system in the on-state for a slightly longer time. The faster fluctuations will also lead to the system turning to the on-state slightly more often. The combination of these two effects leads to the enhanced transmission through the cavity seen for values of $\langle n \rangle_{pump}$ below or in the bistability region. At very high $\langle n \rangle_{pump}$ the opposite effect takes place. For large enough $n_{pol}$ the polariton mode energy shifts above the pump light field energy and the spectral overlap reduces again. The cavity acts as an optical limiter. For slow fluctuations, the cavity transmission will again follow $n_{pump}$ almost instantaneously when it falls below $n_{off}$, while for fast fluctuations, the cavity transmission will return to higher transmission only after a delay as discussed above. In this regime, the delay results in reduced overall transmission.
\begin{figure}[t]
\includegraphics[width=\linewidth]{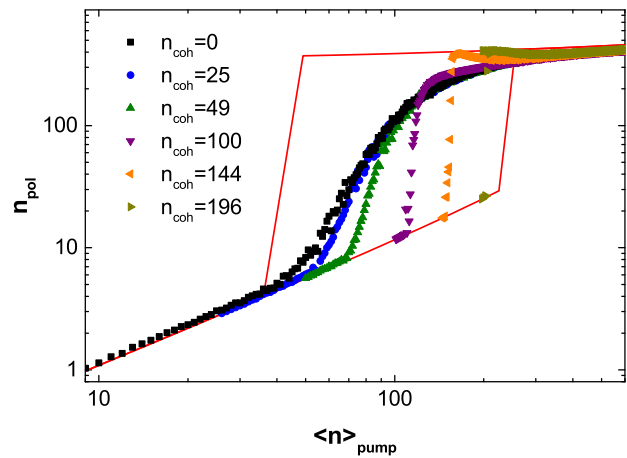}
\caption{(Color online) Mean polariton occupation for pumping with several displaced thermal states in the slow fluctuation limit $\gamma_c \tau_c=15.8$. Red lines show the bistability curves under coherent pumping for comparison. The non-linear rise becomes steeper with increased coherent offset. For large coherent offsets, additional thermal pumping may actually reduce the occupation number.} \label{fig:CoherentOffset}
\end{figure}
After investigating the limiting cases of pure coherent and thermal pumping, it is also worthwhile to study the intermediate case of displaced thermal states, which are equivalent to a thermal field offset by a coherent amplitude. In practice, such a light field may be easily realized by splitting a coherent beam, creating a pseudothermal beam using one of the split beams and recombining the two beams afterwards. The transmission of the polariton system for displaced thermal states with several coherent displacement amplitudes is shown in figure \ref{fig:CoherentOffset}. Red lines denote the results for purely coherent pumping for comparison. For all displacements, a distinctive non-linear increase occurs, which is shifted to larger mean total pump photon numbers compared to pure thermal excitation with increasing coherent fraction and also becomes steeper. Far above $n_{on}$ all curves show the same behavior already identified for pure thermal pumping. While the transmission curve lies in between the two branches of the bistability curve for most pump fields including some thermal component, for large coherent fractions, the transmission indeed reaches the upper branch of the bistability curve directly after crossing the threshold.
\begin{figure}[t]
\includegraphics[width=0.8\linewidth]{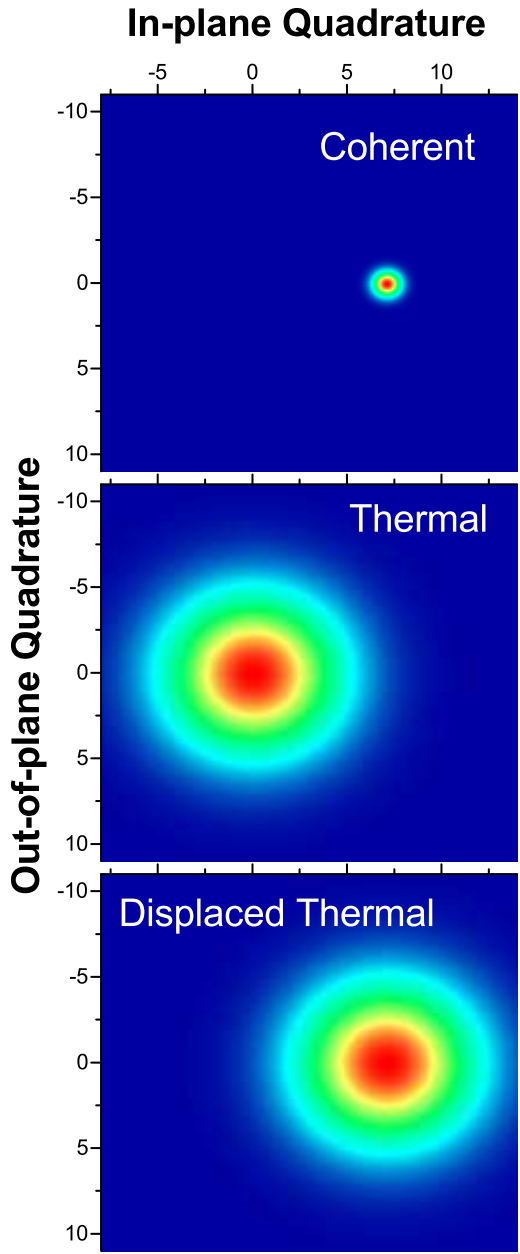}
\caption{(Color online) Wigner quasi-probability distributions of a coherent state with $\langle n \rangle$=100 (upper panel), a thermal state with $\langle n \rangle$=20 (middle panel) and the combined displaced thermal state with $\langle n \rangle$=120 (bottom panel). Note that the displaced thermal state yields significant probabilities for lower field quadratures than the pure coherent state although it has a larger mean photon number.} \label{fig:Wigner}
\end{figure}
It is especially remarkable that the transmission curve lies below the coherent transmission curve for large coherent offset and these results do not depend on whether one starts with a cavity in the on- or off-state. In this regime, the coherent component on its own without any thermal addition would be sufficient to keep the system in the on-state. Surprisingly, adding additional thermal intensity may actually return the cavity to the off-state. This surprising result becomes understandable when taking into account that a displaced thermal state is not simply an incoherent superposition of light from two sources, but a coherent addition of its fields. This can be seen intuitively by looking at the Wigner function of the field, which is a quasi-probability distribution of the quadratures of the light field as shown in figure \ref{fig:Wigner}. As explained earlier, a thermal field field can be described in terms of a random walk of independent oscillators, which is equivalent to a Gaussian distribution in phase space centered at zero with a width given by the mean photon number. The Wigner function of a coherent state is a minimal uncertainty state displaced by a coherent amplitude $\alpha$. A displaced thermal state is a combination of both: A Gaussian distribution centered at some finite displacement. Comparing the Wigner functions of a coherent state and a displaced thermal state of the same displacement, it is immediately obvious that although the mean photon number of the displaced thermal state is higher than that of the coherent state, there is also a finite probability that the instantaneous photon number is much smaller than in the coherent case at some moment. During these periods of low $n_{pump}$, the system will return to the off-state for a short while, which results in the counterintuitive result, that introducing some additional thermal component to a coherent pump beam may actually result in switching a previously stable system off. As a possible application, it might be feasible to control bistability and switch between the two branches of the bistability curve just by adding thermal light.

\section{Polariton Number Statistics}
When investigating the response of a polariton system to the photon number variance of the pump beam, it is straightforward to assume that also the polariton number statistics will show non-trivial properties. Photon number statistics have already been suggested as indicators in order to identify interesting properties like chaos in microlasers \cite{Albert2011}, photon blockade \cite{Gerace2014}, higher-order antibunching \cite{Stevens2014} or bundled photon emission \cite{Muoz2014}. In polariton systems, occupation number statistics have already been utilized to monitor scattering processes\cite{Schmutzler2014}, identification of the polariton laser regime \cite{Tempel2012,Horikiri2010,Assmann2011} and have been proposed as a good indicator for phenomena like analogues to black holes \cite{Gerace2012,Solnyshkov2011}. Generally speaking, it is expected that the noise of $n_{pol}$ should be larger than that of $n_{pump}$ as the non-linearity acts in a similar fashion as an amplifier and phase-insensitive amplification increases noise. We quantify the noise in terms of the equal-time second-order correlation function
\begin{equation}
g^{(2)}(0)=\frac{\langle: n^2_{pol} :\rangle}{\langle n \rangle^2_{pol}},
\end{equation}
where the double stops denote normal ordering of the underlying field operators.
\begin{figure}[t]
\includegraphics[width=0.95\linewidth]{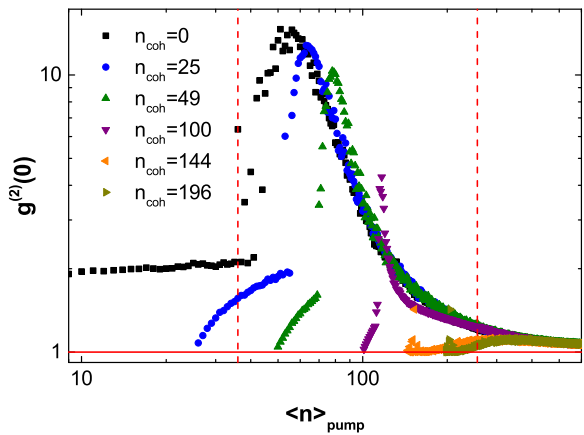}
\caption{(Color online) Equal-time second-order correlation function $g^{(2)}(0)$ of $n_{pol}$ for pumping with several displaced thermal states. The red line denotes the photon number statistics of a purely coherent pump beam for comparison. Dashed lines mark the range of the bistable region for coherent pumping.} \label{fig:g20_offset}
\end{figure}
Figure \ref{fig:g20_offset} displays $g^{(2)}(0)$ for pumping with several displaced thermal states. Below the bistability region the polariton statistics reproduce those of the driving field, which corresponds to values of $g^{(2)}(0)$ between 1 and 2, depending on the relative fractions of thermal and coherent light. A pronounced noise overshoot of up to $g^{(2)}(0)$=14 appears in the bistability region. With increasing coherent displacement, the magnitude of noise decreases and the region of increased noise narrows. These results mirror the effects already seen in the polariton occupation number. With increasing coherent displacement, the non-linear threshold spans a smaller region of photon numbers and it is only this region, which gives rise to the huge polariton number variance. Beyond the threshold, $g^{(2)}(0)$ gradually retuns to lower values again with the notable exception of the two largest coherent displacements. For these two cases, the polariton statistics return to the coherent value of 1 directly behind the threshold before increasing again slightly. Comparing the results with figure \ref{fig:CoherentOffset} directly shows that this sudden appearance of coherence coincides with the transmission moving to the upper branch of the bistability curve. The interpretation of this effect is straightforward. As already discussed, adding a thermal component to a coherent light field yields some probability to increase the instantaneous photon number, but it yields also a smaller probability to reduce it. As shown in figure \ref{fig:Wigner}, this depends on the width of the thermal light field in phase space and therefore on the photon number of the thermal light field. Adding more thermal photons increases the mean photon number of the displaced thermal state, but it also reduces the smallest instantaneous photon number that can occur with reasonable probability. Accordingly both the pinning of the transmission to the upper branch of the bistability curve and the emergence of a coherent polariton state result from the same effect. The bare coherent pump state is too weak to switch the polariton system to the on-state. Adding a small thermal component to the pump yields a reasonable probability for instantaneous photon numbers sufficient to switch to the upper branch of the bistability curve, but the probability for decreasing the instantaneous pump photon number below $n_{off}$ is essentially zero. The system becomes pinned to the on-state.\\

\section{Conclusions}
To conclude, we have theoretically studied the response of a polariton fluid to a noisy driving field showing thermal photon number statistics and to displaced thermal pump states. We found a pronounced dependence of the polariton occupation number on the displacement and the correlation time of the external light field and were able to demonstrate that adding more photons on average may actually lead to a reduced photon number during some instants. We investigated the impact of this effect on a bistable polariton system. Its response shows a large photon number variance overshoot in the threshold region. Such photon extrabunching has attracted considerable attention recently in a wide variety of systems\cite{Boitier2011,Bromberg2014,Iskhakov2012,Hong2012} and may prove useful for multi-photon excitation spectroscopy \cite{Jechow2013} and ghost imaging \cite{Chan2009}. Also, it might be possible to utilize added thermal components for switching in bistable systems. Adding feedback to the system may result in chaotic behavior, which is of interest for optical random number generation\cite{Solnyshkov2009}.
\begin{acknowledgments}
We gratefully acknowledge support from the Deutsche Forschungsgemeinschaft through research grant AS 459/1-2 and the Collaborative Research Centre TRR 142.
\end{acknowledgments}


%\bibliography{report.bib}


\end{document}